\documentclass[aps,prb,twocolumn,superscriptaddress]{revtex4}

\usepackage{graphicx}
\usepackage{dcolumn}
\usepackage{bm}
\begin{document}

\preprint{APS/123-QED}

\title{Effects of out-of-plane disorder on the nodal quasiparticle and superconducting gap  in single-layer Bi$_2$Sr$_{1.6}Ln_{0.4}$CuO$_{6+\delta}$ ($Ln$ = La, Nd, Gd)
}

\author{M. Hashimoto}
\affiliation{Department of Physics, University of Tokyo, Hongo, Tokyo 113-0033, Japan}
\author{T. Yoshida}
\affiliation{Department of Physics, University of Tokyo, Hongo, Tokyo 113-0033, Japan}
\author{A. Fujimori}
\affiliation{Department of Physics, University of Tokyo, Hongo, Tokyo 113-0033, Japan}
\author{D.H. Lu}
\affiliation{Department of Physics, Applied Physics, and Stanford Synchrotron Radiation Laboratory, Stanford University, Stanford, California 94305, U.S.A.}
\author{Z.-X. Shen}
\affiliation{Department of Physics, Applied Physics, and Stanford Synchrotron Radiation Laboratory, Stanford University, Stanford, California 94305, U.S.A.}
\author{M. Kubota}
\affiliation{Photon Factory, Institute of Materials Structure Science, High Energy Accelerator Research Organization (KEK), Tsukuba, Ibaraki, 305-0801, Japan}
\author{K. Ono}
\affiliation{Photon Factory, Institute of Materials Structure Science, High Energy Accelerator Research Organization (KEK), Tsukuba, Ibaraki, 305-0801, Japan}
\author{M. Ishikado}
\affiliation{Department of Physics, University of Tokyo, Hongo, Tokyo 113-0033, Japan}
\author{K. Fujita}
\affiliation{Department of Physics, University of Tokyo, Hongo, Tokyo 113-0033, Japan}
\author{S. Uchida}
\affiliation{Department of Physics, University of Tokyo, Hongo, Tokyo 113-0033, Japan}

\date{\today}

\begin{abstract}
How out-of-plane disorder affects the electronic structure has been investigated  for the single-layer cuprates Bi$_2$Sr$_{1.6}$$Ln$$_{0.4}$CuO$_{6+\delta}$ ($Ln$ = La, Nd, Gd) by angle-resolved photoemission spectroscopy.
We have observed that, with increasing disorder, while the Fermi surface shape and band dispersions are not affected, the quasi-particle width increases, the anti-nodal gap is enhanced and the superconducting gap in the nodal region is depressed.
The results indicate that the superconductivity is significantly depressed by out-of-plane disorder through the enhancement of the anti-nodal gap and the depression of the superconducting gap in the nodal region.
 
\end{abstract}


\maketitle

\section{Introduction}
To reveal how the structural disorder affects the electronic structure of high-$T_c$ cuprates is important to understand how the $T_c$ is suppressed by disorder. Inhomogeneity of the electronic structure has been extensively studied by STM/STS \cite{PanImaging00,HudsonInterplay01,McElroyCoincidence05}, and it has been revealed that both disorder inside the CuO$_2$ plane such as Zn substitution and disorder outside the CuO$_2$ plane such as excess oxygen affect the STS spectra significantly. Angle-resolved photoemission (ARPES) studies on Bi$_2$Sr$_2$CaCu$_2$O$_{8+x}$ (Bi2212) \cite{VobornikSpectroscopic99, VobornikAlternative00} have revealed that the disorder introduced by electron irradiation make'" the pseudogap larger, while the superconducting gap in the antinodal region becomes smaller. Disorder effects and doping dependence have been discussed in Bi$_2$Sr$_{2-x}$Bi$_x$CuO$_{6+\delta}$ (Bi-Bi2201) \cite{PanEvolutionCondmat}. Disorder effects in La$_{2-x}$Sr$_x$CuO$_4$ (LSCO) have been studied by ARPES by Zn or Ni substitutional impurities in the CuO$_2$ plane and by Nd or Ba substitutional impurities in the out-of-CuO$_2$ plane \cite{ZhouDual01, VallaGround06}.

\begin{figure}
\begin{center}
\includegraphics[width=0.8\linewidth]{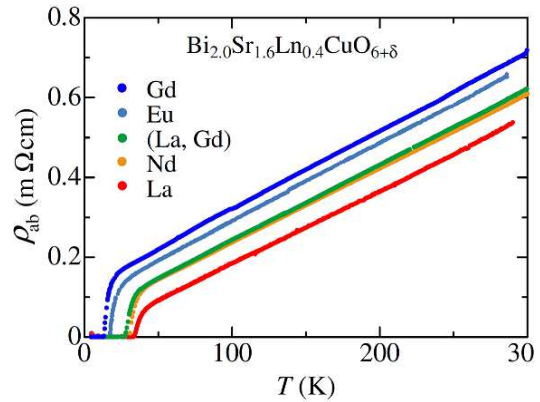}
\caption{(Color online) Temperature dependence of in-plane resistivity
for optimally doped $Ln$-Bi2201 reproduced from Ref. \onlinecite{FujitaEffect05} in which disorder is introduced by changing the $Ln$ species.
}
\label{res}
\end{center}
\end{figure}

The effects of out-of-plane disorder in the high-$T_c$ cuprates have attracted much attention in recent years because it has been demonstrated that the critical temperature ($T_c$) is considerably affected by out-of-plane disorder in spite of the relatively weak increase of the residual resistivity \cite{EisakiEffect04, FujitaEffect05}. 
This is in remarkable contrast with in-plane-disorder such as Zn impurities, which dramatically increase the residual resistivity and act as unitary scatterers \cite{FukuzumiUniversal96}.
Since carriers in the cuprates are induced by the replacement of ions by those with different valences or the addition of excess oxygens in the block layer, all the cuprate superconductors have naturally out-of-plane disorder.
Therefore, it is essential to understand the effects of out-of-plane disorder in order to understand high-$T_c$ superconductors.
Eisaki $et$ $al.$ \cite{EisakiEffect04} have studied the relationship between the $T_c$ and the magnitude of out-of-plane disorder.
They have shown that, as the number of CuO$_2$ planes increases, and the positions of the substituted ions are away from the CuO$_2$ plane, the $T_c$ increases, that is, the $T_c$ increases with decreasing strength of disorder.
A detailed study of the effects of disorder has been reported by Fujita $et$ $al.$ \cite{FujitaEffect05} focusing on Bi$_2$Sr$_{1.6}$$Ln$$_{0.4}$CuO$_{6+\delta}$ ($Ln$-Bi2201, $Ln$ = La, Nd, Eu, Gd), where the degree of disorder is controlled by varying the radius of the $Ln$ ions.
In this system, mismatch in the ionic radius between Sr and $Ln$ causes disorder, and the relationship between the degree of disorder, $T_c$ and the residual resistivity has been revealed.
The in-plane resistivity of $Ln$-Bi2201 \cite{FujitaEffect05} is reproduced in Fig. \ref{res}.

A systematic STM/STS study of out-of-plane disorder effects in $Ln$-Bi2201 by Sugimoto $et$ $al.$ \cite{SugimotoEnhancement06} has shown that, with increasing disorder, the gap size and the gap inhomogeneity increase.
Using ARPES, Okada $et$ $al$. \cite{OkadaOrigin, Okadainfluence} have shown that the pseudogap increases and the Fermi arc length decreases with increasing disorder while the Fermi surface volume does not change. 
The importance of anisotropic electron scattering by impurities in cuprates has been pointed out theoretically \cite{KuliAnisotropic99}.
A recent theoretical study on the transport properties of out-of-plane disorder in cuprates has suggested that the rapid decrease of $T_c$ without strong increase of the residual resistivity can be explained by forward scattering \cite{GraserT[sub07, AbrahamsWhat00}.
Furthermore, to reveal how out-of-plane disorder affects the superconductivity and the anti-nodal (paseudo)gap may give an important clue to understand whether the pseudogap is a precursor of the superconductivity or not, which has been still an open issue \cite{TanakaDistinct06, LeeAbrupt07, WiseCharge08,KanigelEvolution06, NormanDestruction98}.
Therefore, in order to further investigate the effects of out-of-plane disorder, we have studied the electronic structure of the disorder-controlled $Ln$-Bi2201 system by ARPES.
In this paper, the effects of out-of-plane disorder on the Fermi surface, the gap anisotropy and the momentum distribution curve (MDC) width, are presented and discussed.
  
\section{Experiment}
High quality single crystals of optimally doped $Ln$-Bi2201 were grown by the traveling solvent floating zone (TSFZ) method. 
We measured La-Bi2201 ($T_c\sim$34 K), Nd-Bi2201 ($T_c\sim$29 K), La$_{0.2}$Gd$_{0.2}$-Bi2201 ($T_c\sim$27 K), Gd-Bi2201 ($T_c\sim$14 K).
Details of the sample preparation were described elsewhere \cite{FujitaEffect05}.
ARPES measurements were performed at beamline 5-4 of Stanford Synchrotron Radiation Laboratory (SSRL) using a SCIENTA SES-R4000 analyzer with the total energy of $\sim$ 7 meV and the angular resolution of 0.3 $^\circ$. 
Measurements were performed in the angle mode with photon energy $h\nu$ = 19 and 22.7 eV and the polarization angle made 45 $^\circ$ to the Cu-O bond.
The measurement temperature was below $\sim$7 K, well below the $T_c$ of Gd-Bi2201 ($T_c$ $\sim$ 14 K).
The samples were cleaved \textit{in situ} under an ultrahigh vacuum of 10$^{-11}$ Torr to obtain clean surfaces. 
The Fermi edge of gold was used to determine the Fermi level ($E_F$) of the samples and the instrumental resolution. 
ARPES measurements were also performed at beamline 28A of Photon Factory, High Energy Accelerator Research Organization (KEK-PF), using a SCIENTA SES-2002 analyzer with the total energy of 20 meV and the angular resolution of 0.3 degree. 
Measurements were performed with photon energy $h\nu$ = 50 eV.
The polarization angle was 45 $^\circ$ to the Cu-O bond.
The measurement temperature was $\sim$ 9 K.
Samples were cleaved \textit{in situ} under an ultrahigh vacuum of 10$^{-10}$ Torr. 

\begin{figure}
\begin{center}
\includegraphics[width=\linewidth]{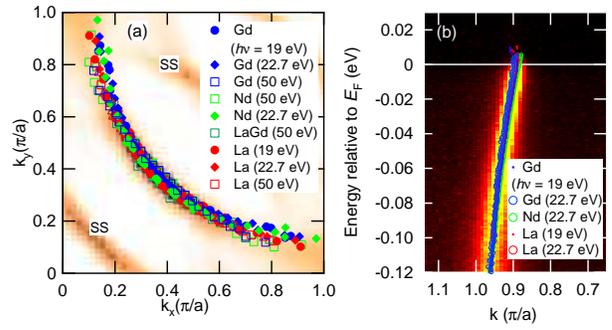}
\caption{(Color online) Fermi-surface shapes and band dispersions in the nodal direction of disorder-controlled Bi2201.
(a) Fermi momentum $k_F$ positions for $Ln$-Bi2201, where $Ln$ = La, Nd, La$_{0.2}$Gd$_{0.2}$ and Gd, and the $k$-space spectral weight mapping for La-Bi2201.
The Fermi momentum $k_F$ positions have been determined from the peak positions in momentum distribution curves (MDC's).
Superstructures due to the Bi-O modulation are denoted by SS.
(b) Band dispersions in the nodal (0,0)-($\pi,\pi$) direction determined from the MDC peak positions.
The differences between different $Ln$'s are within experimental errors.
The energy-momentum ($E-k$) intensity plot is shown for La-Bi2201.
}
\label{map}
\end{center}
\end{figure}

\section{Results and discussion}
\subsection{Fermi surface and anti-nodal gap}
In Fig. \ref{map}(a), we compare the shape of the Fermi surface, namely, Fermi momenta $k_F$ which have been determined from the peak positions of MDC's, for various samples and photon energies.
There is no appreciable difference between the Fermi surfaces of the different samples, that is, out-of-plane disorder does not cause the change of the Fermi surface shape in agreement with Ref. \onlinecite{Okadainfluence}.
The same areas for all the samples of the Fermi surfaces confirm that the doping levels were the same, as the temperature slopes of the resistivity have indicated \cite{FujitaEffect05}.
It has been suggested that the next-nearest neighbor hopping $t'$ may be associated with local structural change by disorder because out-of-plane disorder should affect the apical oxygen positions \cite{FujitaEffect05}.
The present results imply that the average of $t'$ and hence the Fermi surface shape are not appreciably affected by the out-of-plane disorder, although there may be local modulation in $t'$. 
Like the Fermi surface shapes, Fig. \ref{map}(b) shows that one cannot see any differences between the dispersion in the nodal (0, 0)-($\pi$, $\pi$) direction between La-Bi2201 and Gd-2201 within experimental errors.

\begin{figure}
\begin{center}
\includegraphics[width=\linewidth]{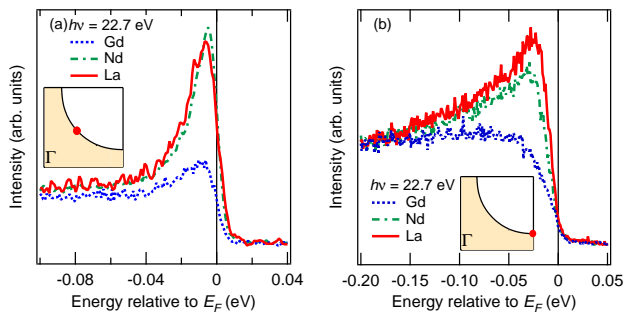}
\caption{(Color online) Energy distribution curves (EDC's) on the (remnant) Fermi surface of $Ln$-Bi2201 samples.
(a) EDC's in the nodal direction.
(b) EDC's in the antinodal region.}
\label{EDC}
\end{center}
\end{figure}

In Fig. \ref{EDC}, we show energy distribution curves (EDC's) on the Fermi surface in the nodal $\sim (\pi/2, \pi/2)$ and the antinodal $\sim$($\pi$, 0) regions for $Ln$-Bi2201 ($Ln$ = Gd, Nd, La) samples.
For the EDC's at the node [Fig. \ref{EDC}(a)], one can see a sharp quasi-particle (QP) peak near $E_F$ but the peak for Gd-Bi2201 is weaker than that for Nd- and La-Bi2201 if the EDC's are normalized at high binding energies ($\sim$-0.2 eV).
As for the EDC's at the antinode [Fig. \ref{EDC}(b)], the spectral intensity from $E_F$ to $\sim$-0.2 eV decreases as the magnitude of disorder increases.
For the most disordered Gd-Bi2201, the spectral intensities dramatically reduced compared to those in less disordered Nd- and La-Bi2201.
This behavior may indicate an enhancement of the anti-nodal gap in the antinodal region by disorder as already indicated by Okada $et$ $al$ \cite{OkadaOrigin}.
Hence, we note that the effects of out-of-plane disorder are stronger in the antinodal region than that in the nodal region, although it is difficult to evaluate the strength of disorder effect quantitatively.
It has been suggested theoretically that the scattering rate in the antinodal region is strongly increased by disorder \cite{KontaniEffect06, GargStrong08, ZhuElastic04, LeePrivate} in good agreement with the present results.
Also the strong disorder effects in the antinodal region are consistent with STM results \cite{SugimotoEnhancement06} where the average of the gap size, which corresponds the antinodal gap in the ARPES spectra, increases with increasing disorder.
The enhancement of the anti-nodal gap size has not been observed for Zn- or Ni-substituted in-plane-disordered high-$T_c$ cuprates \cite{ZabolotnyyEffect06, PanImaging00, HudsonInterplay01, YoshidaPrivate, TerashimaHigh-resolution04}, indicating that the effects of out-of-plane disorder are different from that of in-plane disorder in the antinodal region.

\begin{figure}
\begin{center}
\includegraphics[width=\linewidth]{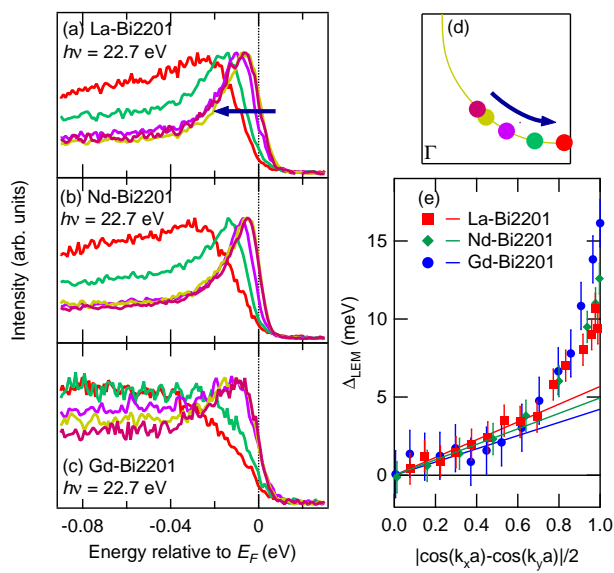}
\caption{(Color online) ARPES spectra and leading-edge mid-point positions along the Fermi surface.
(a)-(c) EDC's along the Fermi surface for $Ln$-Bi2201 ($Ln$ = La, Nd and Gd, respectively).
(d) Circles show the EDC positions for (a)-(c).
(e) Shifts of the leading-edge mid-point (LEM) relative to the node plotted against the $d$-wave order parameter \textbar$ \cos(k_xa)$-$cos(k_ya)$\textbar/2 together with fitted lines. The fitting range is 0 $<$\textbar$ \cos(k_xa)$-$\cos(k_ya)$\textbar/2$<$0.6.
}
\label{gap}
\end{center}
\end{figure}

\subsection{Superconducting gap in the nodal region}
Now, we discuss the depression of the superconducting gap in the nodal region by out-of-plane disorder.
In Fig. \ref{gap}(a)-(c), we have plotted EDC's along the Fermi surface for $Ln$-Bi2201.
We have employed the shift of the leading-edge mid-point ($\Delta _{\mathrm{LEM}}$) as a measure of the magnitude of the gap at each momentum on the Fermi surface and plotted $\Delta _{\mathrm{LEM}}$ against the $d$-wave order parameter [\textbar$ \cos(k_xa)$-$\cos (k_ya)$\textbar/2] in Fig. \ref{gap}(e). 
Here, LEM is defined as the half maximum of EDC.
Using this method, the antinodal gap size and the gap slope are more reliable because they are less affected by the energy resolution, although the absolute gap value in the nodal region where the gap size is small compared to the energy resolution may be difficult to discuss.
One can see that, for La-Bi2201, $\Delta _{\mathrm{LEM}}$ becomes larger from the node to \textbar$ \cos(k_xa)$-$\cos (k_ya)$\textbar/2 $\sim$ 0.7 as usual in a $d$-wave superconductor, and then is deviated toward a larger value for the antinode, in agreement with the previous report \cite{KondoEvidence07}.
In going from La-Bi2201 to Gd-Bi2201, that is, with increasing disorder, $\Delta _{\mathrm{LEM}}$ in the nodal region [\textbar$ \cos(k_xa)$-$\cos (k_ya)$\textbar/2 \textless 0.6] seems to be slightly depressed and hence the gap velocity $v_2$ (velocity along the Fermi surface for the node) slightly decreases.
The gap in the antinodal region, on the other hand, is enhanced in agreement with Ref. \onlinecite{OkadaOrigin}.
It may be interesting to note the similar behaviors in the underdoped Bi2212, that is, the opposite doping dependences between the anti-nodal gap and superconducting gap in the nodal region have been observed \cite{TanakaDistinct06, LeeAbrupt07}. 
The increase of the anti-nodal gap may imply the shrinkage of the Fermi arc as suggested by Okada $et$ $al$. \cite{Okadainfluence}, and therefore decrease $T_c$.
In the present case, however, it is difficult to determine the arc length from measurements below $T_c$.

The depression of the superconducting gap in the nodal region caused by disorder has been predicted theoretically by Haas $et$ $al$ \cite{HaasExtended97}.
Another theoretical study by Dahm $et$ $al$. \cite{DahmNodal05} has indicated that $v_2$ is more strongly renormalized by forward scatterers than unitary scatterers. 
According to Haas $et$ $al$. \cite{HaasExtended97}, the gap in the antinodal region is weakly affected, while we have observed strong enhancement of the anti-nodal gap by disorder.
The tendency of disorder effects in the nodal region observed in the present study is consistent with theoretical predictions \cite{HaasExtended97,DahmNodal05} although the slight decrease of $v_2$ is not enough to explain the large drop of $T_c$.
On the other hand, the anti-nodal gap is enhanced by disorder, contradicting with the theoretical studies \cite{HaasExtended97,DahmNodal05}.
This may indicate the different origins of the anti-nodal gap and the superconducting gap in the nodal region.

\begin{figure}
\begin{center}
\includegraphics[width=\linewidth]{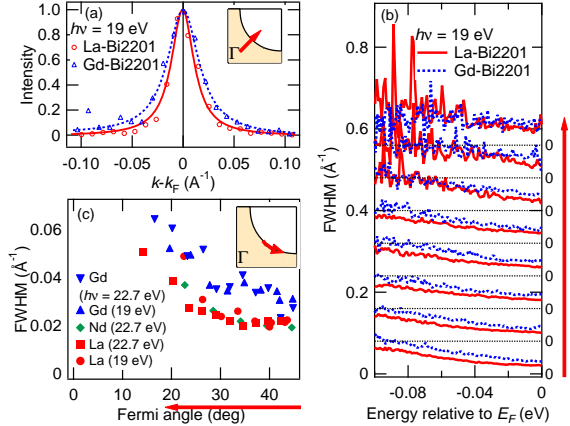}
\caption{(Color online) MDC widths for $Ln$-Bi2201.
(a) MDC's at $E_F$ in the nodal (0, 0)-($\pi,\pi$) direction.
Inset shows the definition of the Fermi angle $\alpha$.
(b) MDC widths determined by a fit to a Lorentzian as functions of energy at various points on the Fermi surface.
The arrow corresponds to that in the inset of panel (c).
(c) MDC widths at $E_F$ as a function of Fermi angle.
}
\label{MDC}
\end{center}
\end{figure}

\subsection{MDC width and transport properties}
In order to see the effects of out-of-plane disorder on the QP mean-free path, we have examined the MDC width as shown in Fig. \ref{MDC}.
We show typical MDC's at $E_F$ in the nodal direction in Fig. \ref{MDC}(a), where one can see that the MDC width is broadened in going from La-Bi2201 to Gd-Bi2201.
The MDC widths as a function of energy for various momenta are plotted in Fig. \ref{MDC}(b), one can see parallel shifts of the MDC width with disorder for all the momenta and energies.
In Fig. \ref{MDC}(c), we have plotted the MDC width at $E_F$ along the Fermi surface to see the momentum dependence of the disorder effects.
The MDC widths again show a constant increase with increasing disorder, indicating that the disorder effect is uniform in this momentum region.
Unfortunately, we were able to obtain reliable MDC widths only around the node because of the slower Fermi velocity ($v_F$) in the antinodal region and the strong influence of the superstructures by the Bi-O modulation.

Very recently, the doping dependence of the MDC width at the node for Bi2212 has been reported \cite{IshizakaDoping-dependence08}.
The increase of MDC width with underdoping has been connected to the anti-nodal gap and the spatial inhomogeneity observed by STM \cite{McElroyCoincidence05}, which is closely related to disorder.
In the present study, because only disorder is controlled while the doping level is fixed, one can unambiguously conclude that the MDC width is clearly affected by disorder.

Now, we discuss the transport properties based on the MDC width in the nodal region, because the transport properties are largely governed by QP's in the nodal region where no pseudogap exists and the Fermi velocity is the highest.
In this region, the difference in the MDC width between the least disordered La-Bi2201 and the most disordered Gd-Bi2201 was $\sim$0.011 $\mathrm{\AA} ^{-1}$, which can be related to the increase of the residual resistivity with disorder.
Thus, we have estimated the increase of the residual resistivity in going from La to Gd from the MDC widths using the Drude formula $\rho=m^*/ne^2\tau=\hbar k_F\Delta k/ne^2$ and compared it with the transport data.
Here, $m^*$ is the effective mass, $n$ is the carrier number, $e$ is the unit charge, $1/\tau$ is the scattering rate and $\Delta k$ is the MDC width.  
Then, the increase of the residual resistivity estimated from ARPES $\Delta \rho ^{\mathrm{ARPES}}$ is found $\sim$2.3 times larger than that from transport $\Delta \rho ^{\mathrm{tr}}$:$\Delta \rho  ^{\mathrm{ARPES}}$/$\Delta \rho  ^{\mathrm{tr}}$ $\sim$ 2.3.
Such a deviation of $\Delta \rho  ^{\mathrm{ARPES}}$/$\Delta \rho  ^{\mathrm{tr}}$ from unity reflects the scattering mechanism caused by the impurities.
Electrical resistivity is largely determined by backward scattering, while the QP lifetime measured by ARPES is determined by all the scattering events.
Therefore, when the scattering is in the unitary limit, $\Delta \rho _0 ^{\mathrm{ARPES}}$/$\Delta \rho _0 ^{\mathrm{tr}}$ = 1, and when forward scattering is dominant, $\Delta \rho _0 ^{\mathrm{ARPES}}$/$\Delta \rho _0 ^{\mathrm{tr}}$ $\gg $ 1. 
The present result $\Delta \rho _0 ^{\mathrm{ARPES}}$/$\Delta \rho _0 ^{\mathrm{tr}}$ $\sim$ 2.3 suggest that out-of-plane disorder act as moderate forward scatterer, as expected from the ratio between the theoretically predicted increase in the residual resistivity in the unitary limit $\Delta \rho _0 ^{\mathrm{unitary}}$ \cite{FukuzumiUniversal96} and the measured increase of the residual resistivity $\Delta \rho _0 ^{\mathrm{tr}}$ ~\cite{FujitaEffect05}: $\Delta \rho _0 ^{\mathrm{unitary}}$/$\Delta \rho _0 ^{\mathrm{tr}}$ $\sim$ 3.7.
These results are consistent with the theoretical conjecture that out-of-plane disorder act as forward scatterer \cite{GraserT[sub07}.

\begin{figure}
\begin{center}
\includegraphics[width=0.8\linewidth]{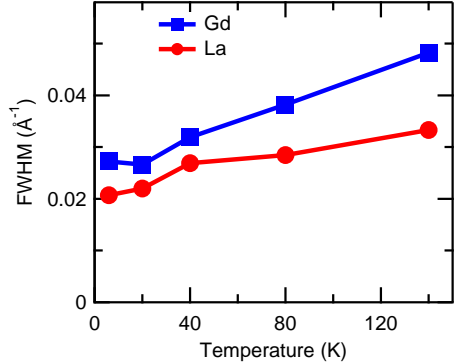}
\caption{(Color online) Temperature dependence of the MDC width at $E_F$ in the nodal direction for La-Bi2201 and Gd-Bi2201.
}
\label{MDCt}
\end{center}
\end{figure}

We have also examined the temperature dependence of the MDC width in the nodal direction as shown in Fig. \ref{MDCt}.
We found that the slope of the temperature dependence of the MDC width is different between La-Bi2201 and Gd-Bi2201, in contrast to the parallel shift of the resistivity with disorder \cite{FujitaEffect05}.
In-plane resistivity mainly detects large-angle scattering therefore the parallel shift of the resistivity means the temperature dependences of the large-angle scattering in La-Bi2201 and Gd-Bi2201 are very close.
On the other hand, ARPES detects all the scattering equally. 
Therefore, the different temperature dependence between La-Bi2201 and Gd-Bi2201 can be attributed to forward scattering.
In Gd-Bi2201, it is suggested that forward scattering is enhanced with temperature compared to that in La-Bi2201, different from large-angle scattering.
Different arc lengths therefore different temperature dependences of the arc length may also be important to understand the discrepancy between the MDC widths and the transport properties.

\begin{figure}
\begin{center}
\includegraphics[width=\linewidth]{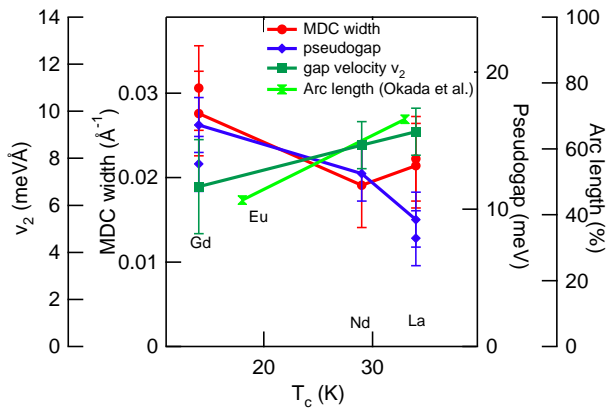}
\caption{(Color online) MDC width at $E_F$ in the nodal direction, anti-nodal gap, arc length (Ref. \onlinecite{Okadainfluence}) and the gap velocity $v_2$ (velocity along the Fermi surface for the node defined in Fig. 4(e)) of $Ln$-Bi2201.
}
\label{sum}
\end{center}
\end{figure}

\subsection{Disorder effects on $T_c$}
In order to discuss the effect of disorder on $T_c$, let us consider the relationship between the superfluid density $\rho _s$, $v_2$, and the effective hole concentration $x$ (which would be proportional to the Fermi arc length \cite{YoshidaLow-energy07}),
\begin{equation}
\hbar ^2 \rho _s /m = \hbar ^2 x /m -2(\ln 2 /\pi \alpha ^2) (v_F / v_2)T,
\label{eq_d}
\end{equation}
proposed by Lee and co-workers \cite{LeeUnusual97, NaveVariational06, WenTheory98}.
Here, $m$ is the carrier effective mass and $\alpha$ is the renormalization factor for the current carried by QP's.
When $\rho $ = 0, $T$ = $T_c$ and therefore Eq. (\ref{eq_d}) leads to
\begin{equation}
T_c = (\hbar ^2 x /m)(\pi/ 2\ln 2)(1 / \alpha^2)(v_2 / v_F).
\label{eq_Tc}
\end{equation}
When $v_2$ becomes smaller for a fixed $x$, $\rho _s$ becomes 0 at lower $T$, which means that $T_c$ decreases with decreasing $v_2$.
As described above, $v_2$ slightly decreases with increasing disorder represented by $\Delta k$ in the nodal region.
The Fermi arc length and hence $x$ may also decrease with disorder according to Okada $et$ $al$ \cite{Okadainfluence}.
Thus, Eq. (\ref{eq_d}) well explains the relationship between the observed disorder effects and the decrease of $T_c$ (-$\Delta T_c$).
In Fig. \ref{sum}, we have plotted the anti-nodal gap, MDC width, $v_2$ and Fermi arc length \cite{Okadainfluence} against $T_c$.
The plot implies that both $v_2$ and the anti-nodal gap are affected by the disorder and affects $T_c$.
Note that the decrease of $v_2$ ($v_{2,\mathrm{Gd}}/v_{2,\mathrm{La}} \sim 0.74$) alone may be insufficient to explain the decrease of $T_c$ ($T_{c,\mathrm{Gd}}/T_{c,\mathrm{La}} \sim 0.41$), and that the increase of the anti-nodal gap and hence the decrease of the Fermi arc length also affects $T_c$ to some extent.

\section{Conclusion}
We have measured the ARPES spectra of disorder-controlled $Ln$-Bi2201 system to investigate the effects of out-of-plane disorder on the electronic structure of high-$T_c$ cuprates.
We have found, a depression of the EDC peak at the node and a broadening of the MDC width $\Delta k$ on the Fermi surface due to moderately forward scattering. 
This may be related with the decrease of the superconducting gap in the nodal region, which can be relatively simply understood as a result of the depression of the $d$-wave superconductivity by disorder. 
In the nodal region, the non-parallel temperature dependence of the MDC width in the nodal direction between Gd-Bi2201 and La-Bi2201 has been also observed, which suggests that a temperature dependence of forward scattering results from the disorder. 
Contrary to the gap in the nodal region, the anti-nodal gap is enhanced by disorder. 
This behavior is opposite to what one would expect for a simple $d$-wave superconductor where the superconducting gap should decrease with disorder, but is consistent with competing order, which would be stabilized by disorder.
We have observed the shape of the Fermi surface and the entire QP dispersion are not influenced by disorder appreciably.
Recently, it has been reported in STM/S studies that there is no $Ln$ dependence in the CDW nesting wave vector $\textbf{q}$ \cite{NoguchiSTM07}, unlike the strongly doping dependent $\textbf{q}$ in Bi2201 \cite{WiseCharge08}.
We suggest that, if CDW exists, without changing $\textbf{q}$, out-of-plane disorder strengthen the CDW order and therefore enhance the anti-nodal gap.
In the antinodal region, the spectral weight depression is observed in a relatively wide energy region ($>$ -0.2 eV).
The bond-centered electronic glass order observed in the wide energy region \cite{KohsakaIntrinsic07, KohsakaHow08}, which can also be stabilized by disorder, is also plausible as the origin of the the spectral weight depression around the antinodal region.
Both the enhancement of the antinodal gap which is probably related to the shrinkage of the Fermi arc and the depression of $v_2$ would reduce the superfluid density and therefore the $T_c$.
We note that, recent ARPES studies reported a simple $d$-wave gap extending to the antinodal region \cite{MengCond} and the co-existence of two energy scales for the antinodal gap \cite{MaCond}, which suggests that further complicated picture of the antinodal electronic structure and its contribution to the superconductivity may be needed to be considered.

\section*{ACKNOWLEDGMENT}
Informative discussion with T.K. Lee is gratefully acknowledged.
This work was supported by a Grant-in-Aid for Scientific Research in Priority Area ``Invention of Anomalous Quantum Materials'' from MEXT, Japan.
This work at SSRL was supported by DOE Office of Basic Energy Science, Division of Materials Science and Engineering, with contract DE-FG03-01ER45929-A001.
The work at KEK-PF was done under the approval of PF Program Advisory Committee (Proposal No.2006S2-001) at the Institute of Material Structure Science, KEK.


\end{document}